# Determinants of the Propensity for Innovation among Entrepreneurs in the Tourism Industry

**Miguel Ángel Montañés-Del-Río * and José Aurelio Medina-Garrido**

INDESS, Universidad de Cádiz, 11406 Jerez de la Frontera, Spain; joseaurelio.medina@uca.es
* Correspondence: miguelangel.montanes@uca.es

**Abstract:** Tourism's increasing share of Gross Domestic Product throughout the world, its impact on employment and its continuous growth justifies the interest it raises amongst entrepreneurs and public authorities. However, this growth coexists with intense competition; as a result of which, tourism companies must continuously innovate in order to survive and grow. This is evident in the diversification of tourism products and destinations, the improvement of business processes and the incorporation of new technologies for intermediation, amongst other examples. This paper expounds on the factors that explain the propensity for innovation amongst tourism entrepreneurs and it may help governments to promote innovation that is based on those determining factors. The hypotheses are tested using a logistic regression on 699 international tourism entrepreneurs, taken from the 2014 Global Adult Population Survey of the Global Entrepreneurship Monitor project. The propensity for innovation amongst tourism entrepreneurs has a statistically significant relationship to gender, age, level of education and informal investments in previous businesses.

## 1. Introduction

Tourism is an economic and social phenomenon because of its ability to stimulate the development of regions, and there is growing recognition of how tourism can be innovative, particularly through creativity. This is evident in the development of new services that integrate technology and sustainable initiatives [1]. Tourism has the capacity to drive forward socio-economic progress [2,3], making it fertile ground upon which to start businesses [4].

In 2015, tourism accounted for 9.8% of the Gross World Product (GWP) and was responsible for 1 in every 11 jobs that were created globally [5]. In 2016, its share of the GWP increased to 10.2% [6]. This rose to 10.4% in 2017, generating 9.9% of the world's employment (more than 313 million jobs), and it is expected to exceed 11% by 2028 [7].

The tourism industry is characterised by its continuous transformation [8], by the saturation of supply and by globalised clients [9], thus creating exceptional and intense competition [10,11]. Therefore, companies in the tourism industry must innovate to achieve lower costs, offering new (or at least, updated) products and services (of a better quality) which would satisfy the requirements of the potential customers [12].

Innovation in tourism is the means with which to grow [13] and survive in the face of the speed and ease with which competitors can copy and implement a successful new idea [14]. This may be an explanation for the wide variety of products and tourist destinations that have come about in response to the demand for new alternative types of tourism [15]. Additionally, innovation is a factor that continuously enhances the experience of the visitor [16] and, given its unique and exclusive character, it is an important driver of economic development, possibly representing a key competitive advantage for tourist organisations [17–21].

A fundamental aspect in the study of the development of the tourism industry is the analysis of tourism entrepreneurs. Although entrepreneurship and innovation are central to the success and development of the tourism industry, there is scant literature on innovative entrepreneurship within the sector [22] and insufficient research on tourism innovation [23,24]. The literature on entrepreneurship in the tourism sector [25–27] has analysed such topics as tourism entrepreneurship [28–30], sustainable tourism entrepreneurship [31–35], innovation in restaurants [36,37] and the relationship between tourism entrepreneurship and innovation [38], but the factors that motivate the more or less innovative behaviour of the entrepreneur in tourism have not yet been studied, hence innovative behaviour remains subject to debate.

Among the innovative initiatives in the tourism sector, there is a growing interest in sustainability, leading to changes in the way these companies operate in the environment [1]. Innovation is pertinent, not only for competitiveness for tourism firms and destinations, but also for enhancing environmental protection and social benefits from the perspective of sustainable tourism [39]. The great importance of sustainability for the competitiveness of tourism destinations worldwide, and especially for developed destinations, has recently been confirmed, which should imply a change in the behaviour of destination managers in relation to sustainability [40]. Since tourism development is mainly based on the exploitation of the natural resources of destinations, the process of tourism innovation should respect the criterion of sustainable development [41]. In the long term, the competitiveness of tourism destinations is linked to their sustainability, i.e., to the protection of natural and cultural resources that guarantee their attractiveness [42]. In fact, sustainability should play an important role, not only in promoting the long-term competitiveness of developing tourism destinations but also, and even more importantly, in the competitiveness of developed destinations [40]. The innovative propensity of a hotel or accommodation service provider, for example, depends not only on the use of technology but also on sustainable initiatives; the latter perhaps being the reason why ecotourism is becoming increasingly popular amongst people who want to connect tourism with the environment and natural settings [1]. Achieving more sustainable tourism is only possible by being more innovative [3]. Thus, eco-innovation in tourism is a path towards sustainable technologies that ensure eco-efficiency and clean production processes [43], such as wine tourism [44,45], confirming the existence of a close relationship between sustainability and tourism, as recognised by the United Nations in its debates on this issue [46].

This paper pursues two objectives: (1) to address the lack of attention that is paid to tourism innovation within political and academic circles, which may be due to the fact that the processes and sources of knowledge for innovation in service companies (such as those in the tourism industry) are either more complex and informal than in industrial companies [13], and (2) to contribute to an increased knowledge of the driving forces behind the propensity for innovation amongst tourism entrepreneurs [14,47].

Discovering the determinants of a tourism entrepreneur's propensity for innovation could cause the proliferation of an ecosystem which, based on certain cultural and social norms that favour innovation (i.e., the public promotion of science and technology [15]), may generate a greater probability that innovative businesses will be created [16]. To achieve this, the present study has a focus on the incidence of sociodemographic and economic variables, and indeed, perceptual and relational variables, and their impact on entrepreneurial innovation in the tourism industry. More specifically, this paper determines the effect that variables which are related to the probability of being an entrepreneur have on the propensity for innovation of those entrepreneurs who create a business in the tourism industry. These variables include age [17–21], the individual's previous level of income [48], gender [49,50], perceptual variables [51,52], variables related to social capital [53–56] and, finally, the explanatory variables of intellectual capital [57–59].

The value of this research is two-fold: (1) it will fill the gap in the literature on innovative touristic entrepreneurship [22,23]; (2) it will produce a profile of the entrepreneur, upon which governments could focus their aid, thus potentially affecting the industry with the greatest potential for innovation, as claimed in the body of literature [60,61], and contributing, through the improvement of its financial results, to the growth and wealth of their country [62].

This paper is divided into five parts. The first part presents the theoretical framework and develops the hypotheses derived from it. The second part explains the design and methodology of this research. The obtained results are reported in the third part, while the discussion is presented in the fourth part. The conclusions are exposed in the fifth and final part of this paper.

## 2. Theoretical Framework and Development of Hypotheses

It is essential to understand what factors drive entrepreneurial behaviour [63]. Some aspects of this behaviour depend on objective sociodemographic variables (e.g., age, gender, educational level or employment status), while others depend on an individual's subjective characteristics (e.g., their perceptions and preferences) [64,65].

In turn, the literature on entrepreneurship emphasises the importance of innovation as a driving force behind the emergence of new companies and, even when that innovation is not radical in nature, companies can produce an industrial sector that alters the competition rules of the market [66]. Thus, innovation is a critical factor in the gaining of competitive advantages in environments that are changing continuously [67].

From the body of literature on entrepreneurship, the variables that are considered to be the determinants of the propensity to become an entrepreneur are used in the present study to discover whether those variables also affect the propensity for innovation amongst entrepreneurs in the tourism industry. Accordingly, following an analysis of determinant variables: sociodemographic and economic variables, perceptual variables, variables related to social capital, and variables related to intellectual capital, several hypotheses are proposed.

### 2.1. Sociodemographic and Economic Variables

Regarding enterprising individuals, previous studies show relationships between their innovation and their gender [68,69]; their age and their intention to innovate [70]; plus their previous income level and their tolerance to uncertainty or innovation [71]. The aim of this study is to determine whether an individual's gender, age and prior income level influence their propensity for innovation when they are starting up a business in the tourism industry.

The literature on the gender of the entrepreneur confirms that gender moderates entrepreneurship [72,73], motives for entrepreneurship [74], and risk-taking in innovation [75]. With regards to the relationship between gender and innovation, previous studies found a weak but positive relationship, confirming that men are more innovative than women [68,69,76,77]. In contrast, in the specific case of the tourism sector, there is evidence of a positive relationship between the culture of change and the female sex [78], which could favour women's innovative potential in this sector.

With regards to age, the previous studies have highlighted a positive relationship between age and entrepreneurship [19,21,79], as age leads to an accumulation of experience and knowledge concerning an industry. Similarly, there is also a positive relationship between the age of entrepreneurs and the radical innovations they can develop [80], due to their greater ability to manage risk through accumulated experience and knowledge.

On another note, the promotion of entrepreneurial creativity requires a minimum income level, given that, below this level, individuals do not have the time, energy or forethought for whatever might be required beyond their own personal subsistence [81]. However, Koellinger [71] determined that, if the economic situation of an individual who is intent on starting up a business is below the average income, or if they are unemployed and have nothing to lose, there is a greater propensity to take a risk and face uncertainty in order to overcome that uncertainty. In that, when an individual's economic situation improves and the available alternatives increase, their need to create a company decreases [82]. This suggests, therefore, that an individual who has nothing to lose and creates a business will be the most innovative due to their greater tolerance of risk and uncertainty [71].

The foregoing arguments enable us to propose the following hypotheses:

**Hypothesis 1 (H1)**. *Being female increases the propensity for innovation as an entrepreneur in the tourism industry.*

**Hypothesis 2 (H2)**. *As an individual's age increases, so does the individual's propensity for innovation amongst entrepreneurs in the tourism industry.*

**Hypothesis 3 (H3)**. *At lower levels of income, there is lesser propensity for innovation amongst entrepreneurs in the tourism industry.*

**Hypothesis 4 (H4)**. *When there are no better employment options, there is greater propensity for innovation amongst entrepreneurs in the tourism industry.*

*2.2. Perceptual Variables*

The study of cognitive factors and what influence they have on entrepreneurship could answer the question of why some individuals, and not others, decide to start their own business [83]. These cognitive factors include perceptual variables, i.e., an individual's beliefs and perceptions [50]. There is a significant correlation between these perceptual variables and the opening of new businesses and they are also related to innovative behaviour amongst entrepreneurs [71]. The perceptual variables of an entrepreneur [83,84] may include: (1) faith in one's own skill set or self-efficacy [85,86]; (2) the fear of failure [71,87]; and (3) the perception of opportunities [71,88].

Regarding self-efficacy, it is noteworthy that the uncertainty an entrepreneur may feel over the success of their endeavour is a significant barrier to entrepreneurship and innovation [89], but this may be reduced if that individual trusts in their own ability to successfully create a start-up and to innovate [90]. A high level of self-confidence is particularly useful for innovative business ideas [71], in which it may not be possible to predict the outcome and for those individuals with no clear and quick feedback [91]. For these reasons, entrepreneurs who innovate will exhibit a higher level of confidence in themselves than those who merely imitate others [71].

Regarding the fear of failure, a reduction in uncertainty, as generated by the perception of self-efficacy, also favours greater risk-taking amongst entrepreneurs [86]. Where others only see risk, an entrepreneur may perceive an opportunity which they feel they are skilled enough to exploit [92]. This predisposition to accepting risk favours the development of innovative products and processes [87]. Innovation is inherently more risky and uncertain than imitation; thus, it could be assumed that innovative entrepreneurs are more likely to accept risk and uncertainty than others who just imitate [71].

Additionally, perceiving untapped opportunities requires an individual to be in a certain state of alertness [93]. Someone who starts a business in order to take advantage of an opportunity they have detected is more innovative than someone who is forced to do so out of necessity [94]. Even though an innovative opportunity may involve doing something new, innovation can also occur if an entrepreneur chooses to make better use of an existing resource [92,95], in order to satisfy new needs or to promote a new business concept [96].

Based on the foregoing, the following hypotheses that specifically apply to the tourism industry can be established:

**Hypothesis 5 (H5)**. *When starting up a business in the tourism industry, believing that one has the necessary skills, knowledge and experience to succeed in doing so increases the propensity for innovation.*

**Hypothesis 6 (H6)**. *Fear of failure decreases the propensity for innovation amongst entrepreneurs in the tourism industry.*

**Hypothesis 7 (H7)**. *When starting up a business in the tourism industry, the perception of nearby good opportunities increases the propensity for innovation.*

*2.3. Variables Related to Social Capital*

Social capital, understood as a set of social relationships that provide resources to help people act effectively [93], facilitates access to useful resources and information for the creation of businesses [97]. The social capital of an entrepreneur can provide access to new knowledge and innovation [95] because an individual's social network can improve their capacity to recognise opportunities and will also inspire new entrepreneurial ideas [52]. Therefore, social capital influences innovative entrepreneurship [56] and product innovation [94], i.e., there is a correlation between social capital and the innovations of entrepreneurs [98]. In addition to the search for social legitimation, the specific features of an entrepreneur's social network, such as meeting other entrepreneurs or having been an informal investor, have an impact on their capacity for innovation [99,100]. In this sense, knowing other entrepreneurs could offer a significant potential for improving innovation [101]. The relationships between entrepreneurs may be determinants of innovative activity in tourism [102], due to the shared knowledge that can be accessed [103–105]. Furthermore, being acquainted with other entrepreneurs has a positive influence on the capacity for innovation because acquaintances have the potential to not only inspire new ideas but they can also encourage the creative combination of innovative ideas [106].

In an industry as globalised as tourism, it can be interesting to create and maintain strong international networks, as they have recently been shown to be a determining factor in improving the export performance of small- and medium-sized enterprises (SMEs) [107].

The pursuit of social legitimation is also an important determinant of entrepreneurship and innovation. Conforming to what a network of entrepreneurs deems as socially appropriate legitimises the created organisation and reinforces the support of its social capital. This, in turn, generates a more favourable environment for gaining access to relevant information and resources, it increases the entrepreneur's tolerance of failure and it enables the entrepreneur to take on more risks [108]. For this reason, the subjective standards of an entrepreneur's social environment allow for social interactions that have a positive relationship with respect to innovation [109,110]. The support an entrepreneur may gain from their social environment can be estimated through indicators such as whether the media reports on entrepreneurial success stories [111] and whether the population believes that being an entrepreneur generates social status and recognition [111,112]. These factors can make entrepreneurship an attractive and prestigious profession that leads to a better standard of living, which is viewed as being gained in a legitimate manner [84].

It may also be relevant to pay special attention to entrepreneurs who were previously considered to be business angels. Their experience as investors and the special social bonds they developed [113] could contribute significantly to increasing their opportunities for innovation [114], as a result of their preference for innovative investment proposals [115].

Based on the foregoing, the following hypotheses for the specific case of the tourism industry can be established:

**Hypothesis 8 (H8)**. *Knowing other entrepreneurs who have started up their businesses in the past two years increases the propensity for innovation amongst entrepreneurs in the tourism industry.*

**Hypothesis 9 (H9)**. *Finding out about the success stories of other entrepreneurs through mass media increases the propensity for innovation amongst entrepreneurs in the tourism industry.*

**Hypothesis 10 (H10)**. *Believing that starting a new business is a desirable career path increases the propensity for innovation amongst entrepreneurs in the tourism industry.*

**Hypothesis 11 (H11)**. *Recognising that successful entrepreneurs are respected and have a high standard of living increases the propensity for innovation amongst entrepreneurs in the tourism industry.*

**Hypothesis 12 (H12)**. *Being a former business angel increases the propensity for innovation amongst entrepreneurs in the tourism industry.*

*2.4. Variables Related to Intellectual Capital*

The intellectual capital of an entrepreneur is related to their level of education, their knowledge of how to start a business, and also any prior entrepreneurial experience [52]. An entrepreneur's intellectual capital influences their ability to extrapolate, interpret and apply information in a way that others cannot [116]. Thus, there is a positive relation between knowledge and the propensity to create a business [117]. With regard to the level of education, it has recently been shown that the higher the level of education, the greater the possibilities for individuals to generate sustainable innovations through the creation of their own business identity [118].

Previous experience of having been an entrepreneur or business angel can have an important impact on an individual's ability to innovate [115], as aforementioned. However, there is also a relationship between intellectual capital that is gained through education and the innovation of an entrepreneur [119,120]. The more highly educated entrepreneurs are more likely to pioneer product innovations [70]. A higher level of education therefore favours the creation and popularisation of innovations [121], because different types of knowledge can be accessed and there is a greater likelihood that such knowledge would be related to potential business opportunities [52]. In fact, there is evidence that the knowledge base of tourism companies may affect innovation on an organisational level [122]. The last hypothesis to be put forward is therefore:

**Hypotheses 13 (H13)**. *The higher the level of studies, the greater the propensity for innovation amongst entrepreneurs in the tourism industry.*

## 3. Methodology

*3.1. Data and Measurements*

The Global Entrepreneurship Monitor (GEM) gathered data from the Global Adult Population Survey (APS) 2014, which were used in the present study to prepare the sample. The APS questionnaire was administered to at least 2000 adults in each of the 73 participating world economies, covering all regions of the world (Africa, Latin America and Caribbean, Asia and Oceania, Europe and North America), representing almost 72.5% of the world's population [111]. Amongst all the questionnaires, those referring to businesses older than 42 months were omitted, this being a common measure in the GEM project to account for early-stage entrepreneurship [64].

There is abundant literature relating entrepreneurship and innovation [62,70,123–126], but to achieve the objectives that are pursued in the present study, a classification was needed that distinguishes between different intensities of innovation amongst entrepreneurs. This allows the creation of a variable called Innovative Tourism Entrepreneur for this work, which will also serve to check the effects on it of factors that are typical of entrepreneurship. Thus, to measure the degree of innovation of new projects, the recommendations of Koellinger [71] and Fuentelsaz and Montero [127] were followed. Koellinger [71] distinguishes between two types of entrepreneurs: imitators and innovators, basing this on the answers given to the questions indicated in Table 1. An imitative entrepreneur who responds with the codes T3, C3 and M1 is referred to as a "pure imitator". In all other instances, the entrepreneur will be deemed to be innovative (to a greater or lesser degree) and is referred to as an "innovator (of any type)". An initial sample of 10,569 pure imitators and 14,241 innovators was obtained using this criterion.

Variable "TEAISIC4_4D" of the 2014 Global APS enables the identification of the ISIC (International Standard Industrial Classification of All Economic Activities, Rev.4; Statistical Division of the United Nations). This was used so that the only cases that were selected came from under the following headings: accommodation, food service, passenger transport, travel agencies and tour operators. With this second criterion, the sample consisted of 351 imitators and 348 innovators, with the value 0 being assigned if the entrepreneur is a "pure imitator" and with value 1 assigned if the entrepreneur is an "innovator (of any type)". The independent variables (Table 2) are grouped into four categories, following the recommendations of Ramos-Rodríguez et al. [84].

**Table 1.** Questions from the GEM (Global Entrepreneurship Monitor) Global APS (Adult Population Survey) on innovation.

| Questions | Possible Answers and Codes |
|---|---|
| For how long have the technologies or procedures been available? | T1: Less than 1 year<br>T2: Between 1 and 5 years<br>T3: More than 5 years |
| How many potential consumers would consider the product to be new or unfamiliar? | C1: All<br>C2: Some<br>C3: None |
| How many businesses offer the same products? | M1: Many<br>M2: Few<br>M3: None |

Source: Adapted from Koellinger [71].

**Table 2.** Independent variables included in the study.

| Category | GEM Variables | Description |
|---|---|---|
| Sociodemographic and economic | GENDER | Gender of the individual interviewed |
| | AGE | Age of the surveyed individual |
| | GEMHHINC | Annual income level (divided into thirds by the GEM) |
| | SUREASON | Reason behind the business venture |
| Perceptual | SUSKILL | Awareness of the necessary knowledge to start a business (self-efficacy) |
| | FEARFAIL | Fear of failure |
| | OPPORT | Perception of good business opportunities in the next 6 months |
| Intellectual capital | GEMEDUC | Level of education, aligned with GEM standards |
| Social capital | KNOWENT | Knowing another entrepreneur who started their own business in the previous two years |
| | NBMEDIA | Finding out about success stories through the media |
| | NBGOODC | Creating a new business is a good career path |
| | NBSTATUS | Recognition that those who successfully start a new business have high social standing and are well respected |
| | BUSANG | Provided the funds for the opening of a new business in the last three years (without taking shares or investment funds into account) |

The majority of these variables are dichotomous (see Table 3), and those which are not (age, income and educational level) were subsequently dichotomised.

**Table 3.** Measurement of independent variables.

| | Measurement | |
|---|---|---|
| GEM Variables | Value 0 | Value 1 |
| GENDER | Male | Female |
| AGE | Up to 44 years old | Over 44 years old |
| GEMHHINC | Low or high income level | Average income level |
| SUREASON | Entrepreneur by opportunity | Entrepreneur by necessity |
| SUSKILL | No self-efficacy | Self-efficacy |
| FEARFAIL | Acceptance of failure | Aversion to failure |
| OPPORT | Existence of opportunities | Absence of opportunities |
| GEMEDUC | Secondary education | Higher education |
| KNOWENT | Does not know other entrepreneurs | Knows other entrepreneurs |
| NBMEDIA | Has not heard about success stories | Has heard about success stories |
| NBGOODC | Entrepreneurship is not attractive | Entrepreneurship is attractive |

| NBSTATUS | Entrepreneurship does not generate social status | Entrepreneurship generates social status |
|---|---|---|
| BUSANG | No prior informal investments | Was previously an informal investor |

*3.2. Analysis*

In the first stage, a bivariate analysis was applied in order to find relationships between each independent variable and the Innovative Tourism Entrepreneur (ITE) variable, thereby obtaining a logistic regression model with a statistically significant predictive capacity [128].

Then, contingency tables were prepared using the Chi-squared test of independence to evaluate the hypothesis of the existence of a relationship between two categorical variables. The counts of categorical answers between two (or more) independent groups were taken [129], calculating the coefficient of correlation and independence [130,131].

In a second stage, and following the works of Arenius and Minniti [50], Minola, Criaco and Obschonka [132] and Fuentes-Fuentes, Bojica, Ruiz-Arroyo and Welter [133], five logistic regression models were designed (see the goodness of fit of the model tests in Table 4; Table 5). In Model 1, the connection between sociodemographic and economic variables and the ITE variable was tested, including the interaction between gender and the rest of the variables in the model. In Model 2, the group of perceptual variables was added to the variables used in Model 1. In Model 3, the connection between the perceptual variables and the ITE variable was verified. Model 4 studied the possible relationship between the set of social and intellectual capital variables and the ITE variable. Lastly, Model 5 tested the possible connections between the variables of Models 2 and 4 and the ITE variable.

Table 4. Goodness-of-fit indicators per logistic regression model.

| Indicators | Model 1 | Model 2 | Model 3 | Model 4 | Model 5 |
|---|---|---|---|---|---|
| −2LL (Log Likehood) | 391.048 [a] | 364.430 [b] | 881.942 [b] | 648.610 [b] | 248.104 [a] |
| Cox and Snell $R^2$ | 0.027 | 0.036 | 0.006 | 0.021 | 0.087 |
| Nagelkerke $R^2$ | 0.036 | 0.048 | 0.008 | 0.029 | 0.119 |

[a] The estimate has ended at iteration number 4 because the parameter estimates have changed by less than 0.001. [b] The estimate has ended at iteration number 3 because the parameter estimates have changed by less than 0.001.

Table 5. Hosmer-Lemeshow test per logistic regression model.

| Indicators | Model 1 | Model 2 | Model 3 | Model 4 | Model 5 |
|---|---|---|---|---|---|
| Chi-squared | 13.874 | 13.874 | 13.874 | 13.874 | 13.874 |
| Degrees of freedom | 6 | 7 | 4 | 8 | 8 |
| Statistical significance | 0.031 | 0.019 | 0.756 | 0.613 | 0.558 |

**4. Results**

Table 6 shows the results of the logistic regression analysis. It can be noted that no empirical evidence has been found of the relationship between the propensity for innovation amongst entrepreneurs in the tourism industry and the sociodemographic and economic variables (Model 1), nor between the propensity for innovation and the perceptual variables (Model 3). However, gender and age do have an effect on the propensity for innovation amongst entrepreneurs in the tourism industry (Model 2), as do "prior positive" informal investments and the individual's level of education (Model 4). In contrast, the perception that entrepreneurship is going to lead to social recognition has a negative effect on the propensity for innovation amongst entrepreneurs (Model 5).

Table 6. Regression coefficients (β) per logistic regression model.

| Variables | Model 1 | Model 2 | Model 3 | Model 4 | Model 5 |
|---|---|---|---|---|---|
| Age | 0.279 | 0.551 * | | | 0.490 |
| Gender | 0.639 | 0.464 * | | | 0.312 |
| Income level | 0.127 | 0.044 | | | 0.133 |
| Motive for entrepreneurship | −0.115 | −0.293 | | | −0.421 |
| Gender AND age | 0.639 | | | | |
| Gender AND income | −0.009 | | | | |
| Gender AND motive for entrepreneurship | −0.351 | | | | |
| Fear of failure | | 0.088 | 0.263 | | 0.171 |
| Perception of opportunities | | 0.207 | −0.193 | | 0.058 |
| Self-efficacy | | −0.512 | 0.064 | | −0.283 |
| Knowing other entrepreneurs | | | | −0.298 | −0.332 |
| Knowledge from the media | | | | −0.035 | 0.024 |
| Desirable option | | | | −0.028 | 0.099 |
| Social status | | | | −0.078 | −0.621 * |
| Education level | | | | 0.458 ** | 0.736 ** |
| Informal investor | | | | 0.528 * | 1.035 ** |

Significant variables (* $p < 0.10$) (** $p < 0.05$).

## 5. Discussion

The growth of the tourism industry and its contribution to the GWP make it an engine of progress [2]. However, it is a highly competitive sector [10,11] and this forces it to innovate continuously in order to survive [12,134]. Due to the importance of innovation for the tourism industry, the present study has sought to identify the factors that influence the propensity to innovate of those who undertake to become entrepreneurs in that sector. This objective is in line with the work of Liu and Cheng [47], which explores the driving forces of innovation in micro-, small-, and medium-sized enterprises (MSMEs) in this industry, in order to understand how they help locate and use resources to increase the success of innovation and sustainable development.

The results obtained in the present study must be confronted with the convulsive moments that have been experienced due to the COVID-19 pandemic, as the tourism industry will be forced to introduce innovative changes that, even with uncertain results, will serve to reactivate demand and guarantee a safe environment for its customers. It is not the purpose of the present study to determine the type of innovation to be implemented when undertaking tourism, but technology can be a good ally at this time because technology is recognised as one of the leading driving forces behind innovation. This conception, even with nuances, can also be applied to tourism, despite the fact that most of its technological development comes from outside, leading to imitative innovation [135]. In fact, tourism service providers are increasingly incorporating technological innovations in both the design and the development of their services. The investments of the tourism sector in communication systems [1] are generating innovation in their products, processes and organisation [135]. Moreover, the literature on technological innovation in the tourism sector adapts the well-known concept of Industry 4.0, establishing the concept of Tourism 4.0 [136], a new ecosystem which is based on the paradigm of the production of high-tech services [137]. Typical 4.0 industry technologies such as the Internet of Things (IoT), Big Data, Artificial Intelligence (AI), Virtual Reality (VR) or Augmented Reality (AR) can help to unlock the innovative potential of the tourism sector [138].

Innovation is vital in tourism for maintaining competitiveness and ensuring the best visitor experience [139]. It is a critical factor in increasing the value added of tourism services and improving business performance [23]. For this reason, a greater understanding of the determinants of innovation within tourism would be useful for the pertinent literature and for enhancing business practices [140]. The results of this study have diverse theoretical implications that can reduce the current gap in the literature on innovative entrepreneurship in the tourism industry [22,141]. On the one hand, the results of this study prove that being a woman is a determining factor in entrepreneurial innovation in the tourism industry, despite the existence of results to the contrary in the literature [68,69,76,77,133,142]. Additionally, the positive relation between age and the creation of a company, a detail which is supported by the previous literature [19–21], is also transferable to the tourism entrepreneur's propensity for innovation. Likewise, it has become clear that business angels are not only more entrepreneurial than the rest of the population [52], but they also have a greater propensity for innovation in the specific case of the tourism industry.

On the other hand, the results show that whilst an entrepreneur's prior level of income is an important determinant of entrepreneurship [71,143], it is not a determining factor in innovation in the tourism industry. Finally, the results show that the perceptions of the entrepreneur, which are useful for predicting the probability of starting a business [50,144,145], do not explain why innovation occurs in business ventures in the tourism industry.

## 6. Conclusions

The findings of this study may also have certain practical implications for the promotion of innovative tourism entrepreneurship by public administrations. The support for entrepreneurialism should be particularly directed at women, people aged over 44 years, those with a high level of education and former business angels, and this aim should be without undermining the support of other groups. Given the importance of higher education in innovative entrepreneurship within tourism, universities that offer courses related to tourism management should include training in entrepreneurship in their study programmes. Universities should also pay special attention to training that addresses the possible innovation strategies that can be developed when creating a business in the tourism industry. Regarding former business angels, value must be placed on the experience they gain from their active participation in other businesses, on the potential transfer of innovations between different sectors, and also on the social networks which they develop in this entrepreneurial environment.

Observing trends and the potential for innovation in the tourism sector could have important managerial implications. Innovative business models could be selected based on the sustainability of the tourism industry or on environmental management practices. The global increase in competitiveness in the tourism industry could allow the implementation of eco-innovations as an element of differentiation between the destinations and the companies that make up the industry, with quality management and sustainable development being increasingly important [146]. Another source of potential innovation could be the use of technologies that are specific to Industry 4.0 such as intelligent devices, that can help us learn about the travel experience of tourists, which are already utilised by hotels [147], Augmented Reality applied to mobile phones [148] so that tourists can take virtual tours of the destinations, "intelligent" hotels to replace personal contacts [149] (in order to protect them from COVID-19 infection, for example), or Big Data to analyse sustainable tourism experiences in the destinations [150]. However, since a lack of innovation implies a lack of adaptation to change, and when changes in the environment and in demand are as drastic as those that are required to adapt tourism businesses to the COVID-19 pandemic, the lack of adaptation and innovation puts the survival of a company at serious risk. The general lack of substantial innovation in the tourism industry is an opportunity that can be seized by the most innovative of tourism entrepreneurs.

This study has some limitations that must be taken into consideration. As non-experimental research, it accepts that control over the variables is less than exhaustive, with the result that it is difficult to separate the effects of all the variables involved [131]. For this reason, various logistic regression models were tested using blocks of variables, as per previous studies [50,151–153]. However, other combinations and numbers of variables could be applied to these models.

Additionally, the fact that the study used data from the GEM project conditioned the number and types of variables that were included in the models, thus restricting the analysis of intellectual capital to the level of education variable, for example. This limitation was accepted due to the advantage of being able to use the large amount of data on entrepreneurs that is publically available to researchers through the GEM project. Notwithstanding, after filtering out the entrepreneurs who are related to tourism, the final sample was not large enough to allow comparisons between different tourist activities. However, this would have been difficult in any case, given that the ISIC codes do not clearly identify the type of tourism company in question. The sample size also prevented a comparison between countries, which would have been interesting, given the cultural and economic differences in the tourism industry and the importance it is given in different places.

This work could lead to future lines of research. In relation to the aforementioned difficulty in drawing comparisons between countries and between tourist activities, it would be interesting to replicate this study with an increased sample size. It could also be revealing to compare the results of this study with other studies on innovation within other industries [14], such as banking [154], sport, culture and entertainment [155], etc. The objective would be to find parallels and to observe similar patterns of behaviour. In turn, it would be interesting to use a model of structural equations so as to be able to examine the combined interaction between the variables that determine the exogenous variable "being an innovative entrepreneur in tourism", as well as to detect the existence of mediating and moderating variables in the model. It would also be worth conducting a qualitative study on the underlying causes of the influence that being female, being older than 44 years, having completed a higher education and having experience as a business angel have on innovative entrepreneurship in tourism. These causes could be diverse or even erratic, and a qualitative study would allow this to be analysed more easily. Furthermore, it is not known exactly what type of innovation the innovative entrepreneurs may have implemented or whether such implementation was a success. With this in mind, it would be valuable to also study the success of the "pure imitator" entrepreneur in the tourism sector. Being an imitator is not necessarily negative, as there is a reduction in uncertainty regarding the problems involving large high-risk investments and in the market's rejection of innovation, and also, losses appear to be more acceptable. Moreover, an entrepreneur may start from an imitative situation but, over time, consider more innovative behaviour. The search for possible determinants of entrepreneurial innovation in the tourism industry should undoubtedly continue, and future studies could incorporate the determinants of innovation that have already been observed in the manufacturing industry, in the influence of IT, in the development of internationalisation strategies, and in the planned [156,157] causal or effectual character [158] of innovation, amongst other factors.

**Author Contributions:** Conceptualization, M.A.M.-D.-R.; data curation, M.A.M.-D.-R.; formal analysis, M.A.M.-D.-R. and J.A.M.-G; investigation, M.A.M.-D.-R. and J.A.M.-G; writing – original draft, M.A.M.-D.-R. Both authors have read and agreed to the published version of the manuscript.

**Funding:** This publication and the research described therein was fully funded by INDESS (University Institute of Research in Sustainable Social Development), Universidad de Cádiz, Spain.

**Conflicts of Interest:** The authors declare no conflict of interest. The funders had no role in the design of the study; in the collection, analyses, or interpretation of data; in the writing of the manuscript, or in the decision to publish the results.